# Shifted rectangular mesh architecture for programmable photonics

Jacek Gosciniak

*Institute of Microelectronics and Optoelectronics, Warsaw University of Technology, Koszykowa 75 st., Warsaw 00-662, Poland*

*Email: jacek.gosciniak@pw.edu.pl*

**Abstract**

Programmable integrated photonics has evolved into a potent platform for implementing diverse optical functions on a single chip through software-driven reconfiguration. At the core of these processors are the photonic waveguide meshes that enable flexible light routing and manipulation. However, recirculating hexagonal waveguide meshes, which currently constitute the basic component of the mesh, are essentially limited by the fixed dimensions of their elementary cells, which consist of as many as six components, limiting their spectral and temporal resolution. These limitations have a detrimental effect on the processing of broadband signals and the application of high-precision delay lines. Here, we introduce the concept of shifted rectangular waveguide mesh architecture for programmable photonics by shifting the adjacent columns or rows by a certain specific value. The operation of these shifted rectangular cells is predicated on the rectangular shape of the cell, which is associated with a smaller number of tunable basic units (TBUs), i.e., four, compared to cells based on a hexagonal mesh, i.e., six. However, at the same time, they allow the signal to be redirected to the input port, which distinguishes them from the regular square mesh-based structure. This approach unlocks new degrees of freedom in programmable photonic circuits, offering enhanced spectral and temporal tunability. Additionally, it paves the way for advanced application in topological photonics, quantum information processing, neuromorphic and high-speed optical computing. The photonic chip arranged in this architecture is capable of implementing one or multiple simultaneous photonics circuits with optical feedback paths and/or linear multiport transformations by the appropriate programming of its resources and the selection of its input and output ports.

**Introduction**

Photonic integrated circuits (PICs) have been recently established as a powerful technology able to manipulate light on-chip with the speed of light [**1-3**]. Compared to integrated electronic circuits that perform digital computations, the photonics circuits carry and process analog information. They may be tailored for various applications for which any specific requirements are required; thus, they are called application-specific photonic integrated circuits (ASPICs) [**1**]**,** making the flow of light essentially fixed and adjusted for this specific application. This requires specific circuit architecture that is designed to implement a specific functionality. Similarly to its electronic counterpart, a major technical breakthrough in photonics technology is expected from a shift from ASPICs to general-purpose photonic programmable devices [**4-9**] that enables precise control of the path and properties of light within a chip through software-controlled actuators [**5, 6, 10**]. In comparison, the general photonic processor enables implementation of several functionalities by suitable programming of a common photonic hardware architecture. It brings additional advantages in terms of low cost and enhanced fabrication reliability when combined with the Generic Integration Model (GIM) and Generic Foundry Model (GFM) [**11**].

Programmable photonic integrated circuits (PICs) represent a new paradigm in photonics that aims to design common integrated optical hardware resource configurations that offer the ability to manipulate the flow of light on a chip during operation and, thus, implement a variety of functionalities through suitable programming. The core of the programmable PIC's processor is a photonic waveguide mesh [**12-14**], a two-dimensional lattice that provides a regular and periodic geometry, formed by replicating unit cells. The possible functions of the circuit and the ways of configurations are determined by its connectivity.

The waveguide meshes may be classified into two main types based on architecture: feed-forward meshes [**6, 15-18**], where the light flows from one side of the mesh to the other, and recirculating meshes [**7, 12, 19-21**], where light may be additionally routed in loops and even back to the input ports. In the case of recirculating meshes, they typically adopt a triangular [**12, 21, 22**], square [**7, 12, 22**], or hexagonal [**12, 14, 22, 23**] geometry, where each side of the cell is formed by a tunable basic unit (TBU) [**5, 12, 23, 24**] composed of a gate [**4, 25**]. The 2×2 gate is the main building block for different mesh architectures that allow for light flows in both the forward and backward directions. Among these, the simplest implementation of a 2×2 analog optical gate is based on a Mach-Zehnder interferometer (MZI) that projects the light from two input waveguides onto two output waveguides [**4, 23**]. It requires at least two adjustable phase shifters to independently control the power splitting $\kappa$ and the relative phase delay $\Delta\phi$. For two fixed couplers in the MZI with 50:50 splitting and coupling ratios, the splitting between output waveguides ranging from 0 % (corresponding to a bar state) to 100 % (corresponding to the cross state) is possible. An alternative 2×2 gate combines a controllable coupler and one phase shifter; thus, similarly to the MZI arrangement, two degrees of freedom are provided [**4**]. These building blocks, meshes, are referred to as programmable unit cells (PUCs) and serve as the core of reconfigurable systems, enabling a light flow in both the forward and backward directions. Many such structures comprise the Field Programmable Photonics Gate Array (FPPGA) [**4, 5, 7**] that is analogous to a Field Programmable Gate Array (FPGA) in electronics, offering a unique opportunity to create a flexible platform with versatile functionalities ranging from optical interconnects (quantum information processing) to microwave photonics (neuromorphic computing) [**26-28**]. And as in electronics, they may be combined with ASPICs, leading to hybrid chips in which a programmable core is embedded inside a custom ASPIC. However, compared to FPGA, it does not perform digital logic operations but rather exploits optical interference to perform very high-speed analog operations that act on the phase and amplitudes of the optical signals.

The FPPGA features a programmable photonic core that is comprised of programmable unit cells (PUCs) and is driven and monitored by RF control and electronics, as well as optical and RF input/output ports [**4, 5, 7, 10**]. Software algorithms and programming layers provide users with access to photonic functionality.

However, the typical achievable length and propagation delay in a unit cell of a programmable photonic waveguide array of an FPPGA are fixed and limit the flexibility in the spectral and time domain responses of the circuits that may be implemented. Therefore, the search for smaller unit cells that are highly programmable and interconnected is of great interest.

**Results**

**Shifted rectangular mesh architecture**

The present results pertain to programmable photonics, which is characterized by photonic waveguide mesh arrangements comprising rectangular cells that are shifted to each other, either vertically or horizontally.

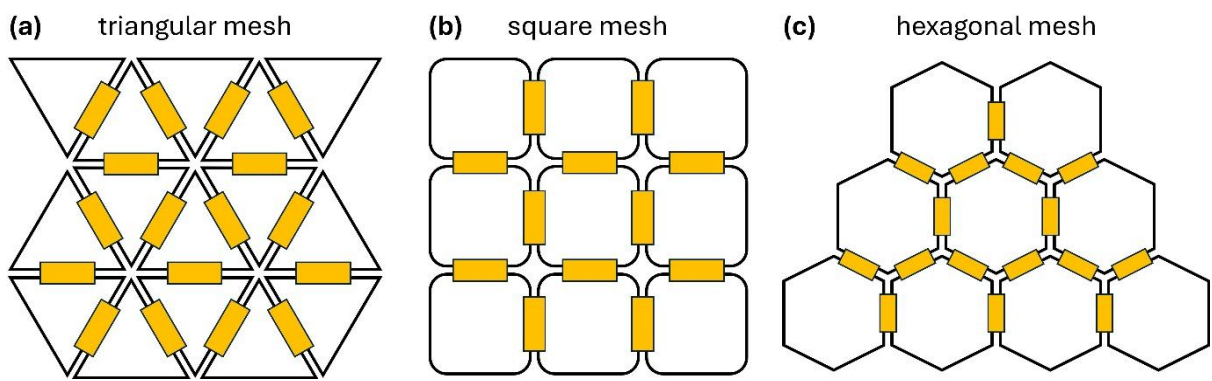

**Figure 1.** State-of-the-art integrated waveguide recirculating mesh architectures: (a) triangular mesh, (b) square mesh, (c) hexagonal mesh.

**Fig. 1** shows the various meshed circuits and segmentation options in tunable basic units (TBUs). They may be divided into three main topologies: (a) triangular meshes, (b) square meshes, and (c) hexagonal meshes. All of these may be discretized into identical tuning units and utilized for programmable photonic integrated circuits.

The starting point of these meshes is a 2D waveguide mesh formed from a replication of a basic tuning element implemented by means of two waveguides coupled by an independent (in power and phase division) tunable basic unit (TBU). The tunable basic unit (TBU) is configured by means of tuning elements based on MEMS, thermo-optic tuning, electro-optical tuning, or optomechanical or electro-capacitive tuning.

The tunable basic unit (TBU) can be preferably implemented by means of balanced or tunable Mach-Zehnder interferometers (MZI), a double actuation directional coupler, or a tunable directional coupler and a phase-shifting element and representable by means of a 2×2 transmission matrix. Depending on the orientation and the interconnection of the TBUs, uniform (square, hexagonal, triangular, etc.) or non-uniform topologies are originated if each TBU has an arbitrary length and orientation.

Standard triangular, rectangular/square, or hexagonal meshes consist of three, four, and six TBUs, respectively, which determine the resonant cavity length and impose constraints on the free spectral range (FSR) and time sampling rate, which limit the processor's applicability in broadband and high-speed operations [**12, 22**].

As the FSR is inversely proportional to the cavity length, as in the case of Ring Resonators (RR) or the length mismatch between two arms of the Mach-Zehnder interferometer (MZI), the smaller FSR is preferred for a realization of both finite impulse response (FIR) and infinite impulse response (IIR) filters with MZI and RR, respectively. Thus, the meshes with smaller TBUs are preferred. On the other hand, the hexagonal mesh architecture is, up to now, the only one where all ports may be used as inputs and outputs. Despite these limitations imposed on rectangular mesh, it is widely used due to its simpler architecture compared to hexagonal mesh and more intuitive programming algorithms.

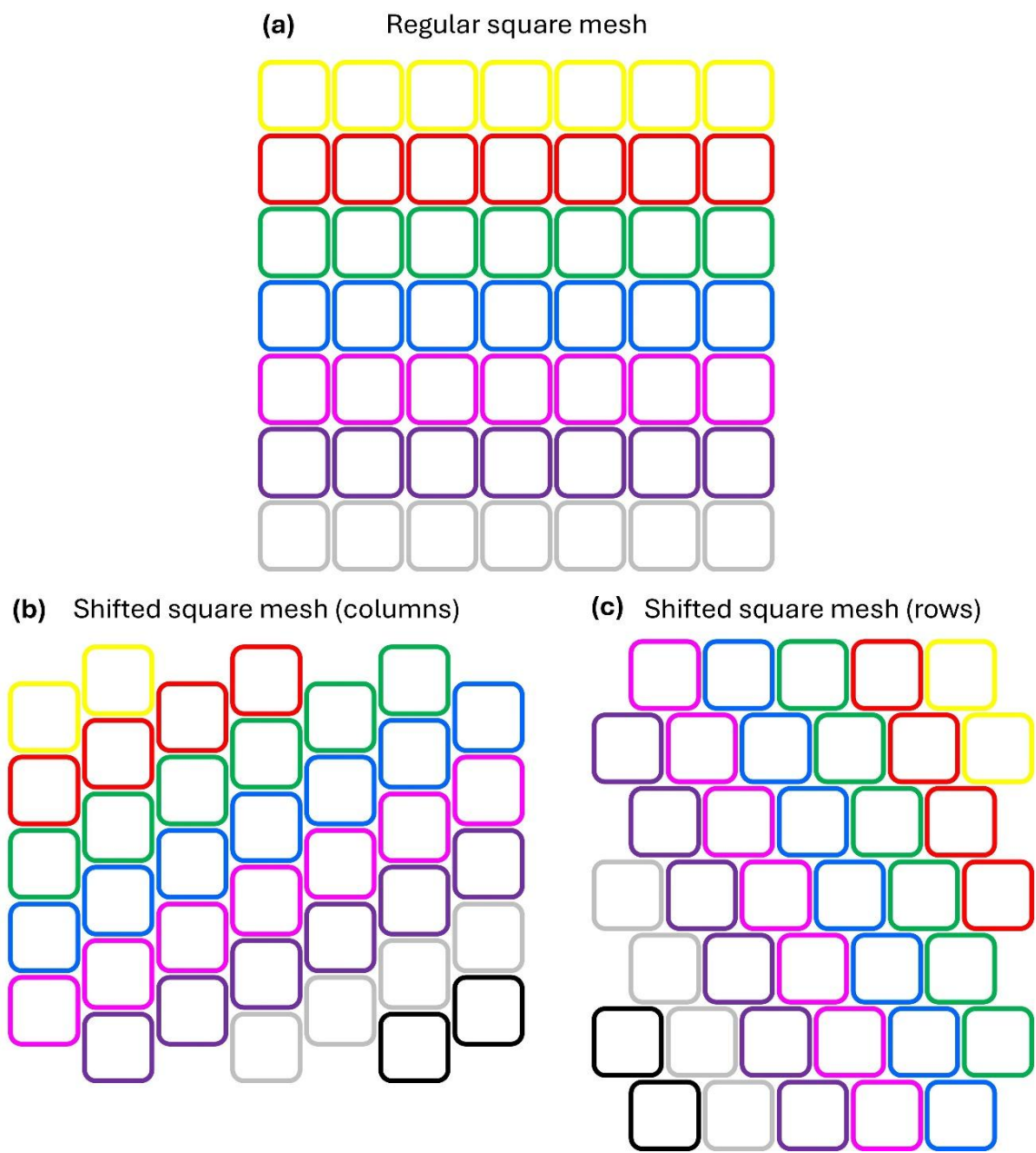

**Figure 2**. (a) Standard square mesh architecture and (b, c) shifted square mesh architecture in columnar (b) and (c) in rows arrangement.

**Fig. 2** shows a standard rectangular/square mesh architecture and a new approach to it, in which the corresponding columns (**Fig. 2b**) or rows (**Fig. 2c**), constructed from the programmable unit cells (PUCs) that consist of tunable basic units (TBUs), are shifted relative to each other by a certain specified value. In the optimal solution, the shift between the corresponding columns or rows is half of the length of a basic unit, PUC, as shown in **Fig. 2b, c**. Thus, the basic unit, PUC, comprises two to four TBUs (see **Figs. 5–7**), depending on the selected configuration. Two TBUs arranged horizontally can function based on a balanced MZI (3 dB MZI), while the other TBUs, which are arranged vertically, can function based on either a tunable coupler with an additional phase shifter, a balanced MZI (see **Fig. 4**), or any other alternative arrangements. This refers to a configuration in which successive columns are shifted relative to each other by a certain value, as shown in **Fig. 2b**. The situation is reversed in a configuration where successive rows are shifted relative to each other (**Fig. 2c**). As a result, the number of TBUs in a single mesh is reduced from six to four, compared to a hexagonal mesh. Consequently, the optical path length is reduced, which decreases the optical insertion losses and enables the synthesis of large FSR filters. Simultaneously, this allows the removal of restrictions on rectangular/square meshes regarding the use of all available ports as inputs and outputs.

Similarly to the hexagonal topology, the proposed shifted rectangular topology yields a more efficient 3-point interconnection scheme compared to a regular rectangular/square topology where four TBUs are interconnected through 4 points.

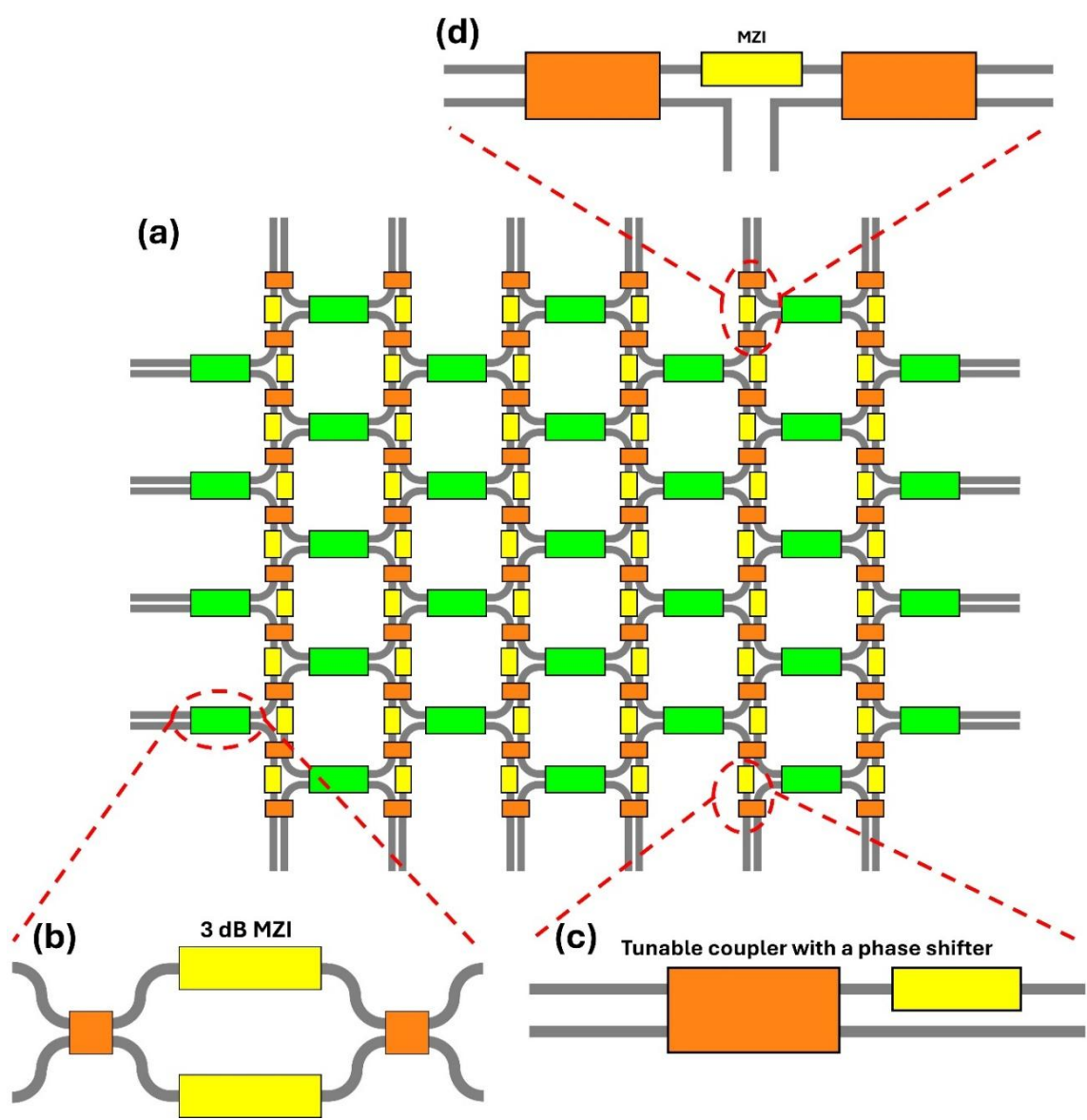

**Figure 3**. (a) The schematic diagram of the programmable photonic processor consisting of the main building blocks – (b) 3 dB (balanced) Mach-Zehnder Interferometers (MZI), (c) tunable coupler with a phase shifter, or (d) tunable Mach-Zehnder Interferometer (MZI).

**Fig. 3a** shows a schematic diagram of the programmable photonic core, which comprises a plurality of rectangular programmable unit cells (PUCs) implemented by way of a series of photonic waveguide elements developed on a photonic chip substrate. In this specific example, the corresponding columns of a rectangular PUC are shifted relative to the previous ones by a certain specified value. Each of these blocks possesses programmable characteristics and can propagate the light in both directions.

**Fig. 3b-d** shows schematics of tunable basic units (TBUs) that can be used as the variable beam splitters in a single rectangular mesh to direct a light through a mesh.

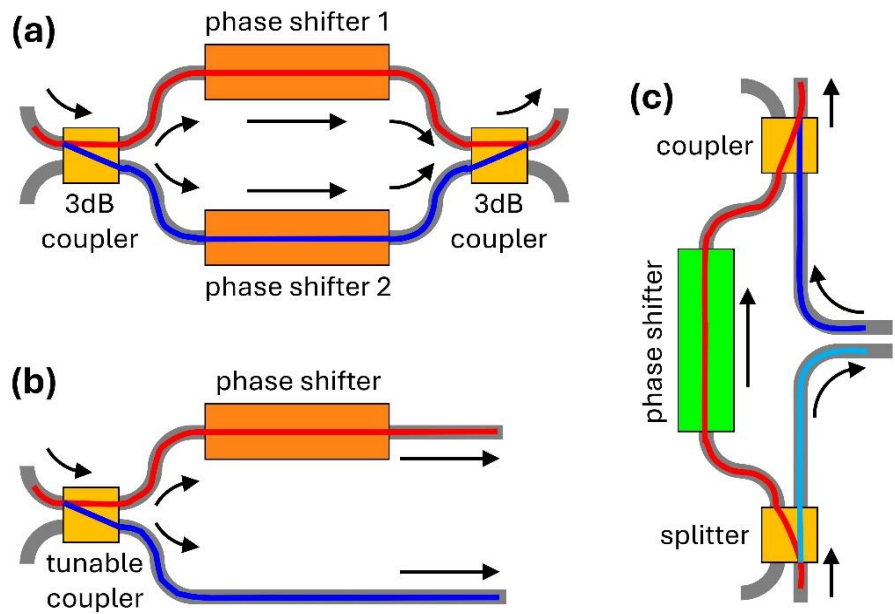

**Figure 4**. The main building blocks of the programmable unit cells: (a) a balanced Mach-Zehnder interferometer, (b) a tunable coupler with a phase-shifting element, and (c) a tunable or balanced Mach-Zehnder interferometer (MZI).

The specifics of the TBUs illustrated in **Fig. 3b-d** are delineated in further detail in **Fig. 4**. As depicted in **Fig. 4a**, the configuration comprises a balanced Mach-Zehnder interferometer comprising two beam splitters and two phase shifters, with each phase shifter positioned at each interferometer's arms [**4, 23**]. Thus, the MZI incorporates two input arms, two delay arms, and two output arms. The input arms, delay arms, and output arms can be constructed from waveguides in order to propagate the optical mode. As illustrated in **Fig. 4a**, the primary shifter is positioned on the top delay arm, while the secondary phase shifter is located on the bottom delay arm.

The two beam splitters can be configured as 50:50 beam splitters and may include directional couplers, multi-mode interferometers, or other beam-splitting mechanisms as are known in the relevant field. The first phase shifter and the second phase shifter, disposed on the top and bottom arms of the MZI, adjust an internal phase $\theta$ and $\phi$ (between 0 and $2\pi$) of light travelling through the arms and can control the coupling ratio of the top and bottom input arm and top and bottom output arm. Therefore, tuning both phases can adjust the splitting ratio and differential output phase, respectively, thereby allowing any beam splitter (i.e., any 2×2 unitary) to be implemented.

The splitting ratio for a balanced MZI loaded with heaters on both arms is obtained by modulating the effective index of waveguides in the upper or/and lower arm due to the Joule effect, thereby producing a corresponding phase shift $\theta$ (upper) and $\phi$ (lower) in the arms. Simultaneous activation of both heaters will cause both arms of the interferometer to experience simultaneous phase shifts determined by the voltage applied to the heaters, which will enable independent control of both the amplitude ratio and phase at the system output. The transfer matrix for a balanced MZI is defined as:

$$T = \frac{1}{2}\begin{bmatrix} 1 & -i \\ -i & 1 \end{bmatrix}\begin{bmatrix} e^{-i\theta} & 0 \\ 0 & e^{-i\phi} \end{bmatrix}\begin{bmatrix} 1 & -i \\ -i & 1 \end{bmatrix} \qquad (1)$$

where directional couplers are fixed to 50:50 splitters.

Another example of the TBU that is based on a tunable beam splitter is illustrated in **Fig. 4b**. It includes two parallelly placed waveguides located at a short distance from each other that enable the interaction of their evanescent fields. One phase shifter is located on two waveguides in a coupling section that enables active control of the splitting ratio, denoted by κ, between the waveguides. The secondary phase shifter is located on one of the output arms, with the purpose of regulating the relative phase between the two output arms.

The transfer matrix for such a device consists of the coupling matrix of a directional coupler or an MMI (multimode interferometer) and the matrix of a phase delay section placed in one of the output arms [**21, 22, 24**]. Thus, the transfer matrix of a tunable beam splitter with an additional phase shifter is defined as:

$$T = \begin{bmatrix} \sqrt{1-\kappa} & -i\sqrt{\kappa} \\ -i\sqrt{\kappa} & \sqrt{1-\kappa} \end{bmatrix}\begin{bmatrix} e^{-i\phi} & 0 \\ 0 & 1 \end{bmatrix} \qquad (2)$$

where κ is the coupling coefficient for the directional coupler and $\phi$ is the phase shift in one of the output arms. For a fixed length of a coupling section and an operation wavelength, the coupling coefficient depends only on the refractive indices of the waveguide core and cladding materials. The integration of a phase shifter in a coupling section enables the tuning of the effective index difference between the two waveguides, and therefore, the resulting coupling coefficient κ. Thus, the coupling coefficient can be modulated through an applied voltage. When modifying the coupling coefficient by actuating the phase shifter in a coupling section, we can produce a power splitting ratio variation between two output waveguides while an external phase shifter introduces a phase delay to one of these output waveguides [**21**].

On the contrary, **Fig. 4c** shows a modified MZI, similar to the one shown in **Fig. 4a**, where one of the phase shifters was removed from a delay arm and a light from a delay arm was directed to the adjacent rectangular cell. The second beam splitter/coupler collects a light provided through a first delay arm and coming from the adjacent rectangular cell. In consequence, a light interferes at the beam splitter/coupler and may be directed to one of the output arms. The transfer matrix for this device is very similar to the transfer matrix of a balanced MZI where a second phase shift is realized in the adjacent rectangular cell.

The two beam splitters can be 50:50 beam splitters, as in the case of balanced MZI (**Fig. 4a**), or can work as tunable couplers that enable an active control of a splitting ratio κ between waveguides (**Fig. 4b**). The second phase shifter is disposed on one of the delay arms to control the relative phase of the two output arms.

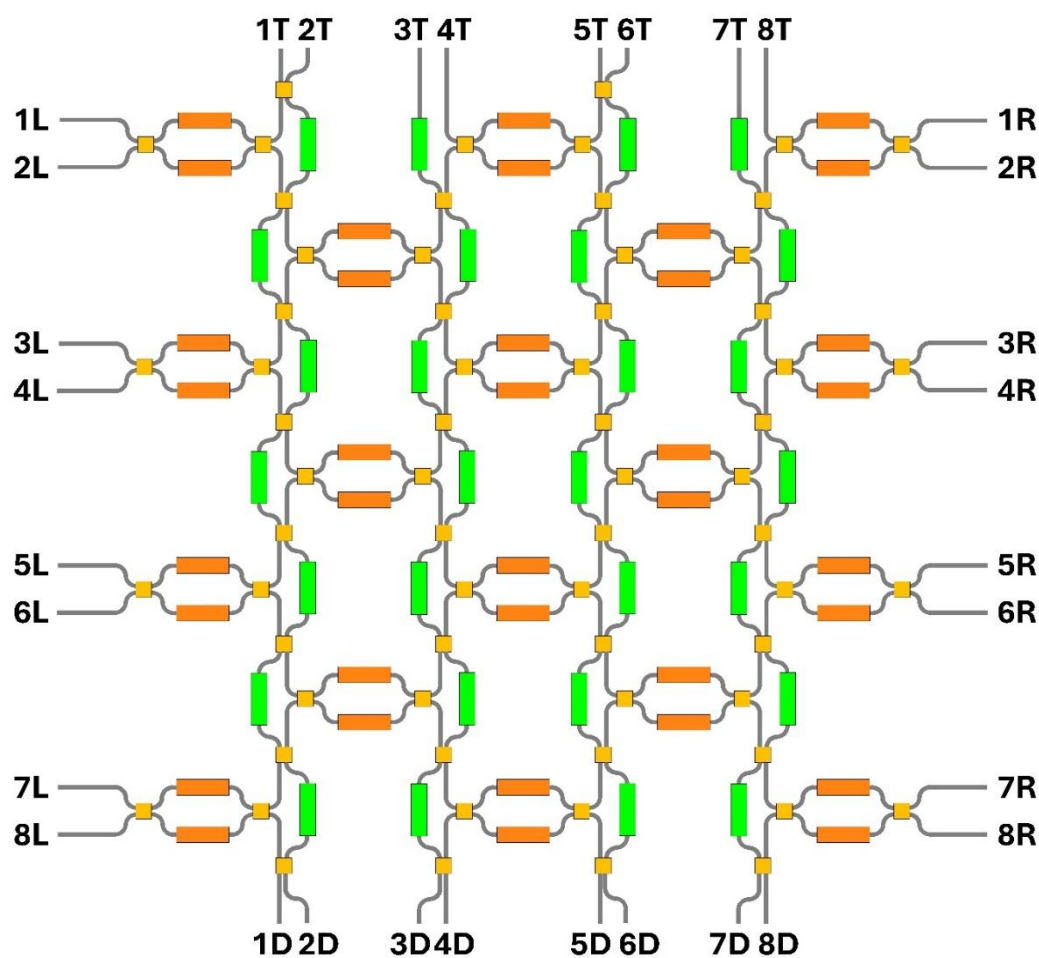

**Figure 5**. An example of one of the proposed shifter rectangular mesh arrangements in which each rectangular cell consists of four tunable basic units.

A more detailed view of a proposed photonic waveguide mesh arrangement is shown in **Fig. 5**, in which each rectangular cell consists of four tunable basic units (TBUs) with corresponding connections between rectangular cells. The presented mesh architecture was realized in a columnar arrangement in which adjacent columns were shifted in a vertical direction for a distance equal to half of the length of the rectangular cell. The connections between rectangular cells are realized through tunable basic units (TBUs).

The main building block in this arrangement is a balanced Mach-Zehnder interferometer (MZI) (**Fig. 4a**) that is organized in a horizontal direction through the entire mesh structure, while the connections in the vertical directions are realized by tunable couplers and phase-shifting elements, as illustrated in **Fig. 4b**.

These vertical elements distinguish the proposed configuration from standard feed-forward network architecture based on MZI, as they enable light to be directed in any direction, including in the opposite direction.

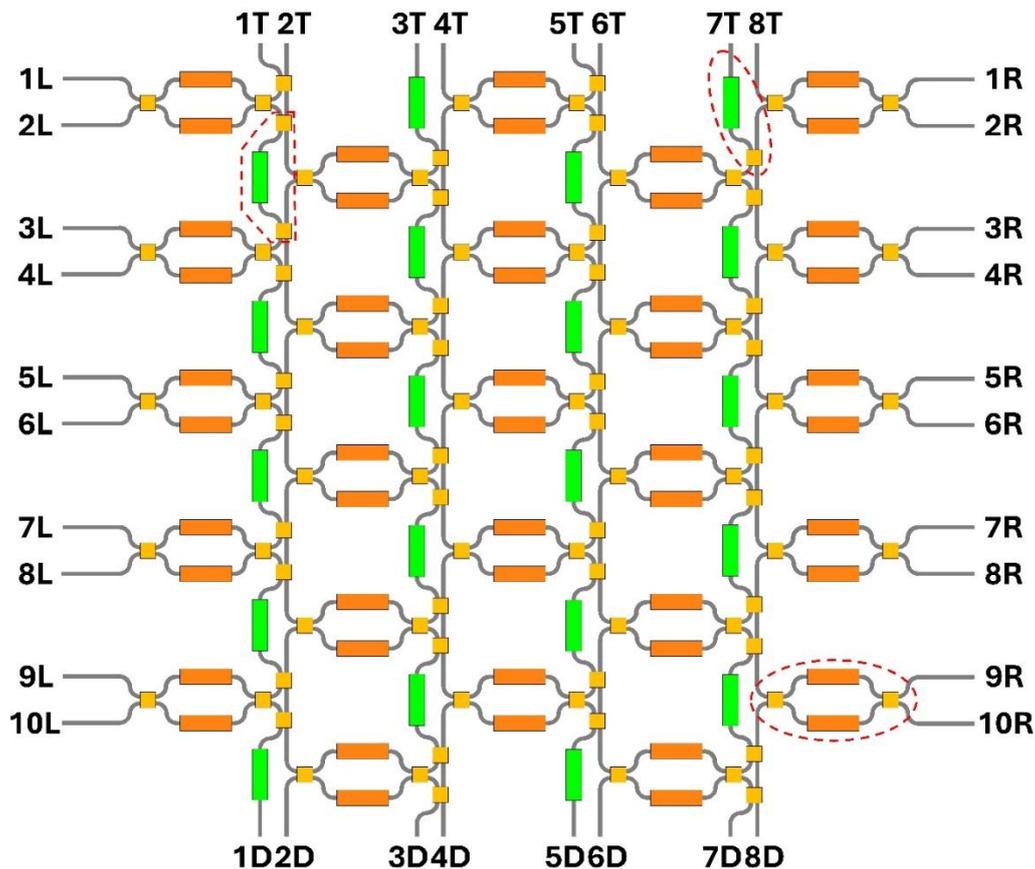

**Figure 6**. An example of one of the proposed arrangements in which each rectangular cell consists of three tunable basic units.

Conversely, **Fig. 6** shows a modification of the proposed photonic waveguide mesh arrangement, in which each rectangular cell consists of three tunable basic units, and the connections between the cells are realized through a balanced Mach–Zehnder interferometer (MZI) in a horizontal direction and a modified MZI in a vertical direction. A columnar arrangement of the presented mesh architecture was achieved by shifting adjacent columns vertically by half the length of the rectangular cell. Any of the ports on all sides of the mesh arrangement defined by numbers 1D, 2D, 3D, etc. can serve as both input and output ports.

Upon comparison of **Fig. 5** with **Fig. 6**, it becomes evident that the unit cell is substantially reduced in size when the unit cell contains a smaller number of TBUs. This facilitates the configuration of systems with an enlarged FSR, thereby diminishing the time sampling rate and consequently enabling operation at a higher velocity.

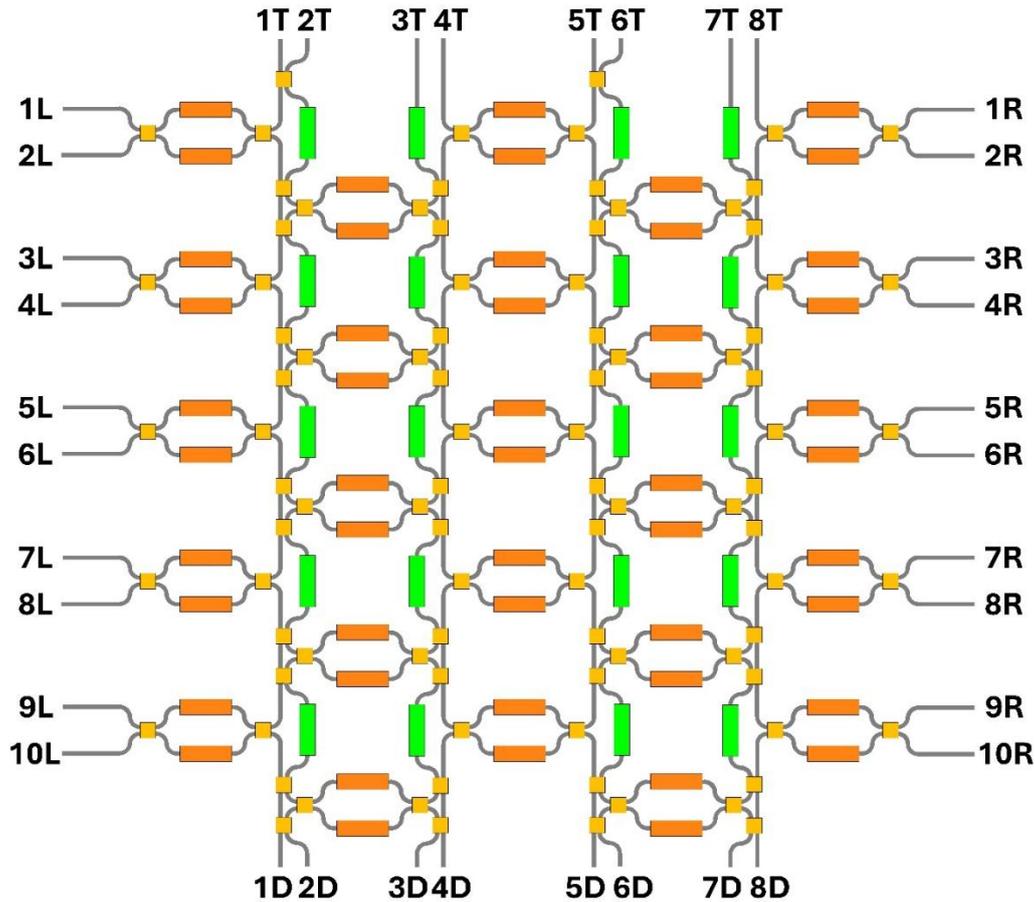

**Figure 7**. An example of one of the proposed arrangements in which each column of rectangular cells with four tunable basic units is connected alternately to a column of rectangular cells with two tunable units.

**Fig. 7** shows a third modification of the proposed photonic waveguide mesh arrangement in which each column of rectangular cells with four tunable basic units is connected alternately to columns of rectangular cells with two tunable units. As previously, a presented mesh architecture was realized in a columnar arrangement in which adjacent columns were shifted in a vertical direction for a distance equal to half of the length of the rectangular cell.

In this arrangement, columns of cells with four TBUs are adjacent to columns of cells with only two TBUs arranged exclusively in the horizontal direction. The size of each cell, both with four and two TBUs, corresponds to the size of a cell with three TBUs, as can be seen from the comparison of **Fig. 5** and **Fig. 7**.

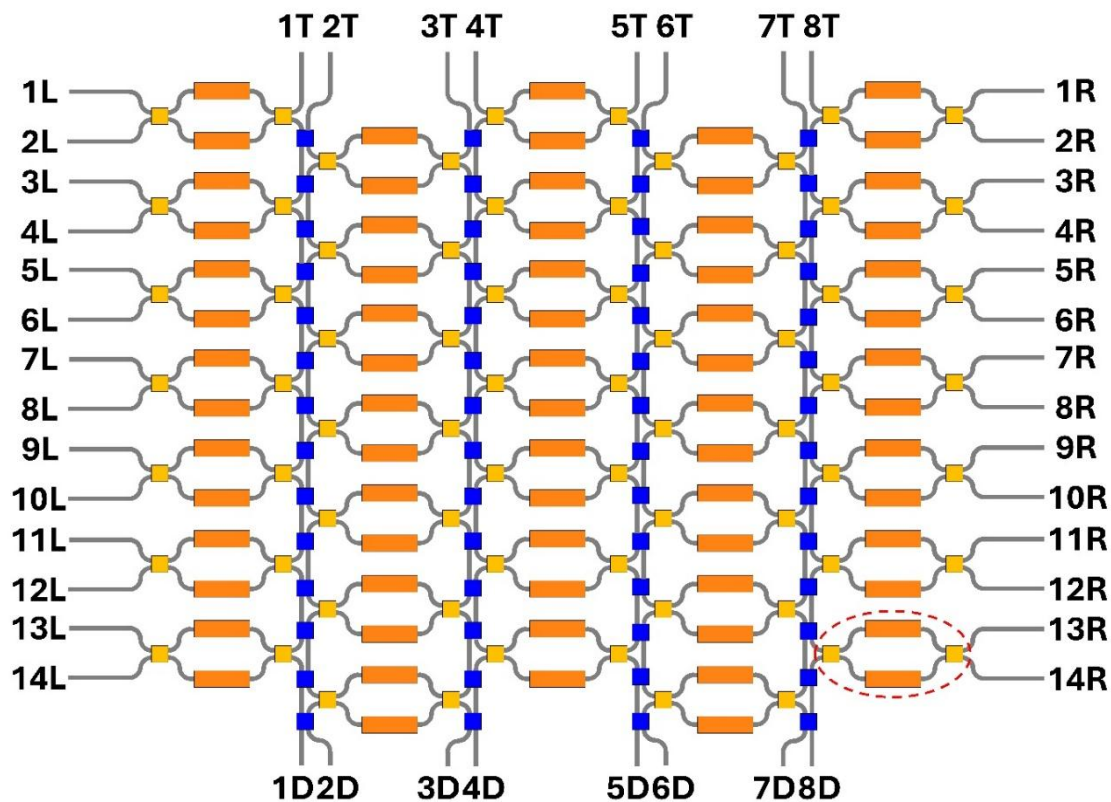

**Figure 8**. An example of one of the proposed preferred arrangements in which each column of rectangular cells with two tunable basic units is connected to a column of rectangular cells with two tunable units.

Another modification of a proposed photonic waveguide mesh arrangement is shown in **Fig. 8,** in which each column of rectangular cells with two tunable basic units arranged in a horizontal direction is connected to another column of rectangular cells with two tunable basic units. The presented mesh architecture was realized in a columnar arrangement in which adjacent columns were shifted in a vertical direction for a distance equal to half of the length of the rectangular cell. As indicated previously, a similar arrangement can be achieved for a row arrangement. Here, the tunable basic unit comprises a balanced Mach-Zehnder interferometer, while vertical connections (in the case of columnar arrangement) are realized through the tunable couplers.

The configuration with two TBUs is very similar to feed-forward mesh architecture, but unlike it, it allows the signal to be redirected back to its initial port or in any other direction. Thus, information flow is not limited to only one direction, e.g., left-right, up-down, but opens up new possibilities in which all ports can be both input and output ports, as is the case with recirculating meshes. This allows for the implementation of control loops and filters based on ring resonators (RRs), which significantly expand the portfolio of feed-forward meshes. This is achieved using tunable couplers arranged vertically, shown as blue squares in **Fig. 8**.

As shown above, the dimensions and sizes of each unit cell are defined by the TBU. The most common on-chip implementation of TBU is based on a balanced Mach-Zehnder Interferometer (MZI) with the phase-shifting elements placed in each arm of the MZI, as shown in **Fig. 4a** [**4, 23**]. By increasing the effective index in one of the arms of the MZI, the phase shift between arms is created, which adjusts the splitting ratio at the termination of the MZI. In consequence, a programmable unit cell (PUC) may work under three operation regimes: cross state, bar state, and tunable state with any possible splitting ratio between both arms of the MZI (**Fig. 10g, h, i**).

An alternative implementation of TUB is based on the tunable coupler (or dual-drive tunable couplers) with a phase-shifting element placed in the arm of the MZI (**Fig. 4b**) [**21, 22**]. The tunable coupler can be realized with a directional coupler that has a variable coupling ratio, which enables a desired splitting ratio between both arms, while a phase-shifting element introduces a phase shift between both arms.

**Implementation of basic photonic components**

The core concept of waveguide meshes is to route light and perform analog matrix and filtering operations through software-controlled manners. Among these, programmable wavelength filtering is essential for the manipulation of signals in the optical domain and stands out as one of the most prevalent and widely adopted functionalities [**29-31**].

Spectral filters are very important components used in nearly every optical application spanning telecommunications, environment, and biological sensing. Among these, the integrated spectral filters offer an important step toward miniaturization. However, the traditional filters suffer from the lack of flexibility, as they cannot change the spectral profile of the device once fabricated. On the contrary, programmable filters enable a single hardware platform to implement a wide range of spectral responses utilizing a number of mesh architectures [**12, 22**].

The proposed programmable shifted rectangular mesh architecture can implement classical FIR and IIR discrete-time impulse filters [**29**]. In the case of FIR filters, both transversal and lattice filters can be realized. Lattice filters, which are usually based on Unbalanced Mach-Zehnder interferometers (UMZIs) and are 2-input/2-output periodic notch filters, find applications as linear phase filters, multi-channel selectors, group delay compensators, etc. To meet the requirements, they have to maintain a desired path length difference *ΔL* between arms by suitably tuning each TBU within the mesh.

In comparison, the IIR filters are implemented through ring resonators, which are either 1-input/1-output or 2-input/2-output periodic filters. In this way, they can implement IIR notch filters and bandpass filters [**29**] that are the basic building blocks for more complex filter designs such as, for example, CROWs and SCISSORs. Those filters are described by the cavity length *ΔL* that is defined by the size of a unit cell. Thus, the smaller the unit cell, the higher the FSR and, in consequence, the higher the operation bandwidth.

Beyond filtering structures based on FIR and IIR, the proposed shifted rectangular waveguide meshes also have to implement programmable delay lines and multiport interferometers [**32**]. The delay lines can be programmed by configuring some TBUs as tunable couplers and some as optical crossbar switches (**Fig. 10g, h, i**), creating adjacent light-path with an incremental length value *ΔL* that is expressed in discrete values of BULs.

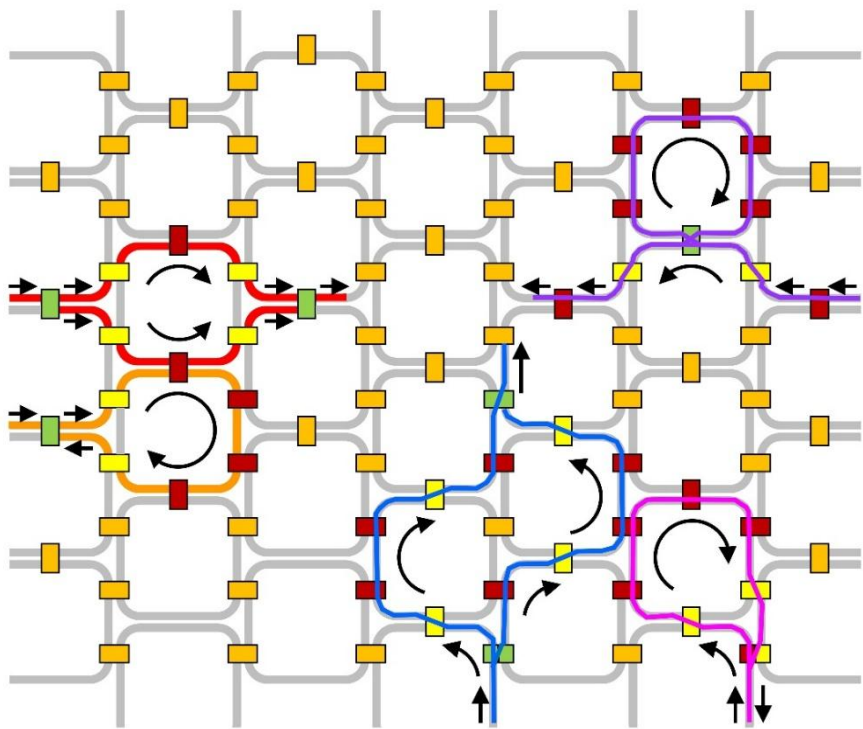

**Figure 9**. A schematic representation of the proposed shifted rectangular mesh architecture with a realization of some basic photonic components such as, for example, balanced Mach-Zehnder Interferometers (MZI), ring resonators (RR), etc.

The BUL, basic unit length, is an important parameter that defines the size of the unit cell. The basic unit length (BUL) of the cell in any mesh configuration is mostly limited by the length of the MZI, which is around $L_s$=450 µm, which corresponds to a propagation delay time $t_s$ of 6.3 ps [**14**]. The length of the MZI is the main factor determining the size of the mesh; thus, the smaller the length of the MZI, the smaller a unit cell. For hexagonal meshes where the minimum cavity length is $L=6L_s$, it corresponds to a round trip time of 40 ps that results in a spectral period in the range of 25 GHz. Compared to it, for any rectangular/square meshes, the smallest cavity length is reduced to $L=4L_s$, which corresponds to a round trip time of 26.7 ps, resulting in a spectral period of 38 GHz. For a shifted rectangular/square mesh, higher cavities follow in increments of 2 BULs. In comparison, for a hexagonal mesh, the second cavity length is 10 BULs, while a higher cavity appears with a step of 2 BULs [**12, 22**].

For an asymmetric MZI waveguide, the lowest length difference between MZI arms is only 2 BULs for a shifted rectangular/square mesh, which is lower compared to both a regular square mesh (4 BULs) and triangular mesh (3 BULs) and similar to a hexagonal mesh evaluated at 2 BULs.

Here, the concept of programming the FPPGAs is demonstrated by six generic designs in **Figs. 9** and **10**. **Figs. 10a-e** show how the configuration of each processing block leads to the programming of two optical filters based on a Mach-Zehnder Interferometer (MZI) (**Fig. 10a, b, d, e**) and a ring resonator (RR) (**Fig. 10c**), while **Fig. 10f** shows how a light may be rerouted back to the input port.

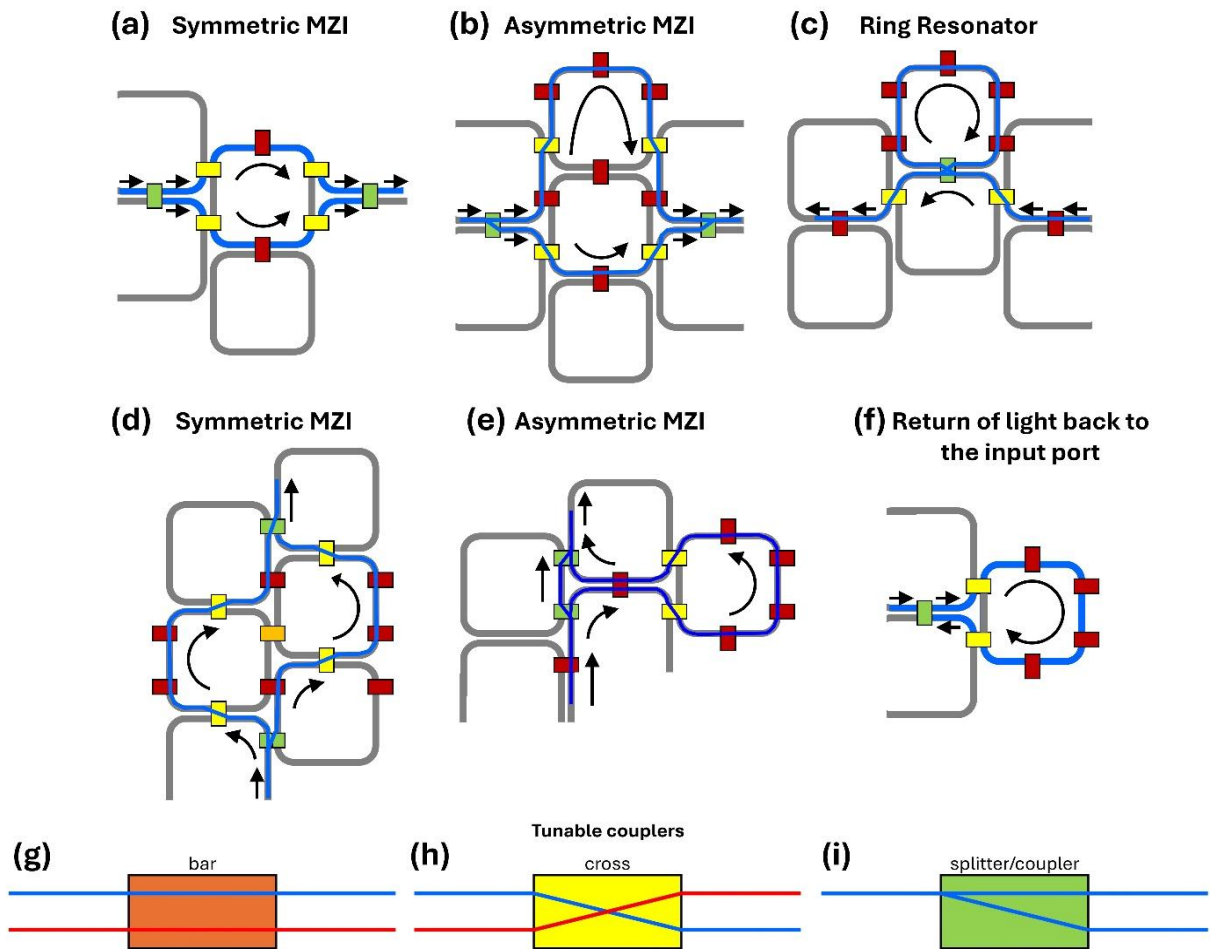

Figure 10. A basic programming of medium-complex circuits in a proposed shifted rectangular waveguide mesh topology realized in a horizontal direction: (a) symmetric (balanced) MZI, (b) asymmetric (unbalanced) MZI, and (c) optical ring resonator; and in a vertical direction: (d) balanced symmetric MZI and (e) asymmetric MZI. (f) Control loop with an input port serving simultaneously as an output port realized in a horizontal direction.

The reconfiguration performance of the mesh, defined as the number of filters with different frequency separation values for the RR filter, shows the best performance for a proposed shifted rectangular mesh, where it was calculated at 11 for a BUL of 25. In comparison, for a hexagonal mesh and triangular mesh for the same BUL of 25, it was calculated at 9 and 6, respectively, while for a regular square mesh, it was 6.

The reconfiguration performance for the MZI filter with a shifted rectangular/square mesh exceeds 12, which is similar to the performance for the hexagonal mesh evaluated at 12. For a triangular mesh, it was calculated at 8, while for a regular square mesh, at 6. As observed from above, the shifted rectangular/square mesh doubles the performance compared to a regular square mesh architecture.

**Monitoring of power in the mesh**

Programmable photonic integrated circuits (PICs) require scaling up hundreds or thousands of optical meshes, which involves precise fabrication of tens of thousands of optical interferometers. However, static component errors induced by process variation grow exponentially as the system expands, which limits the usability of the system. For example, a small beam splitter variation of only 2 % may degrade accuracy by nearly 50 % for feedforward circuits. Thus, hundreds of thousands of optical paths need to be monitored through control of the actuators. It may be realized by monitors or photodetectors, which are key components of the configuring techniques.

Such detectors or monitors should be placed in any part of the photonic mesh to monitor in real time an optical signal in a system. They should operate under low optical power requirements just to sample the small amount of power, which can be convenient for calibrating or setting up the network. Up to now, the state-of-the-art monitoring devices are based on "mostly transparent" detectors [**33, 34**] in the waveguide or can use waveguide "taps" to sample a small amount of power from the guide to a conventional detector [**18**].

An alternative approach relies on an implementation of an in-line Surface Plasmon Detector (SPD) that exploits the photothermal effect to monitor the optical power in a waveguide. It leverages the resistance variation of a plasmonic strip that is in contact with the waveguide and, thus, with a propagating mode. A small portion of the guided optical power is absorbed by the metal stripe and causes a local temperature rise that can be detected by probing the temperature-dependent resistance of the electrode. The responsivity for such a system was experimentally validated at 7.5 µV/mW at a wavelength of 1550 nm [**35, 36**].

The power monitoring with the Wheatstone bridge configuration is another realization of a monitoring device that can be implemented in a photonic mesh [**37, 38**]. It relies on the resistance measurements of the material supporting the photonic or plasmonic mode, where the absorption losses of the propagating mode cause an increase in the material temperature and, in consequence, its resistance. As all elements of the Wheatstone bridge are arranged on the same chip, it allows reducing the influence of environmental temperature fluctuations. The square mesh seems to be the perfect configuration for power monitoring with Wheatstone bridges, as both are based on a square arrangement.

The measurements performed for a monitoring of an optical power propagating in a plasmonic mode revealed very high responsivities reaching up to 6.4 µV/µW for a bias voltage of only 245 mV [35], which is a very promising result and highly exceeds an inline plasmonic detector [38]. However, both monitors suffer from high losses related to plasmonics.

To avoid the inevitable losses associated with plasmonics, the Wheatstone bridges can be based on the transparent conductive oxide (TCO) materials, for example, ITO, AZO, GZO, etc. A thin layer of TCO material can be deposited on top of the waveguide with a separation between the waveguide and TCO provided by a thin oxide layer. The TCO material properties can be chosen or actively modified to be close to the epsilon-near-zero (ENZ) point that ensures enhanced absorption in the TCO [**39-41**]. Furthermore, as it provides a high electron temperature, low thermal conductivity coefficient, and high thermal coefficient of resistance, it can exhibit a significant resistance change under absorption of light; thus, it can be a preferred material for a realization of power monitoring in the photonic circuits.

**Conclusion**

We have proposed a novel mesh design geometry for implementing tunable optical cores in programmable photonic processors based on a shifted rectangular mesh architecture. This design has been analyzed and compared with previously proposed triangular, regular square, and hexagonal mesh topologies, producing very promising results.

This study is an important step forward for programmable photonic meshes. It allows the unit cell size to be reduced and a higher FSR to be achieved compared to hexagonal meshes that are the best available at the moment.

This approach unlocks new degrees of freedom in programmable photonic circuits, offering enhanced spectral and temporal tunability. Additionally, it paves the way for advanced application in topological photonics, quantum information processing, neuromorphic and high-speed optical computing.

**Acknowledgement**

Author acknowledges the constant support of Warsaw University of Technology, Poland for the completion of this work. Furthermore, he is very thankful to Prof. D. G. Misiek for his support and very valuable suggestions.